\documentclass[12pt,preprint]{aastex}

\def\deg{\ifmmode^\circ\else$^\circ$\fi}

\def\kpc{\ifmmode h^{-1}{\rm kpc}\else$h^{-1}{\rm kpc}$\fi}
\def\kms{\ifmmode {\rm km~s}^{-1}\else${\rm km~s}^{-1}$\fi}
\def\bii{\ifmmode b^II\else b$^{II}$\fi}
\def\lii{\ifmmode l^II\else l$^{II}$\fi}
\def\feh{\ifmmode {\rm [Fe/H]}\else [Fe/H]\fi}

\shortauthors{Beers et al.}
\shorttitle{Broadband $UBVR_CI_C$ Photometry of HK Survey and HES Stars}

\begin{document}
\title{Broadband $UBVR_CI_C$ Photometry of Horizontal-Branch and Metal-Poor Candidates
from the HK and Hamburg/ESO Surveys. I.} 

\author{Timothy C. Beers\altaffilmark{1,2,3}}
\affil{Department of Physics \& Astronomy, CSCE: Center for the Study of Cosmic Evolution,
and JINA: Joint Institute for Nuclear Astrophysics, Michigan State University, E.
Lansing, MI 48824}
\email{beers@pa.msu.edu}

\author{Chris Flynn\altaffilmark{1}}
\affil{Tuorla Observatory, Piikki\"o, FIN-21500, Finland}
\email{cflynn@astro.utu.fi}

\author{Silvia Rossi\altaffilmark{1}}
\affil{Instituto de Astronomia,  Geof\'{i}sica e Ci\^{e}ncias Atmosf\'{e}ricas, Departamento de Astronomia, 
Universidade de S\~{a}o Paulo, \\ 
Rua do Mat\~{a}o  1226, 05508-900 S\~{a}o Paulo, Brazil}
\email{rossi@astro.iag.usp.br}

\author{Jesper Sommer-Larsen}
\affil{Dark Cosmology Centre, Niels Bohr Institute, University of Copenhagen,
Juliane Maries Vej 30, DK-2100 Copenhagen, Denmark}
\email{jslarsen@tac.dk}

\author{Ronald Wilhelm}
\affil{Department of Physics, Texas Tech University, Lubbock, TX 79409}
\email{ron.wilhelm@ttu.edu}

\author{Brian Marsteller\altaffilmark{2}, Young Sun Lee\altaffilmark{2}, Nathan
De Lee \altaffilmark{2}, Julie Krugler\altaffilmark{2}}
\affil{Department of Physics \& Astronomy, CSCE: Center for the Study of Cosmic Evolution, and JINA: Joint Institute for Nuclear
Astrophysics, Michigan State University, E. Lansing, MI 48824}
\email{marsteller@pa.msu.edu, leeyou25@msu.edu,delee@pa.msu.edu,
kruglerj@msu.edu}

\author{Constantine P. Deliyannis}
\affil{Department of Astronomy,Indiana University, Swain West 409, 727 East
Third Street, Bloomington, IN 47405}
\email{con@astro.indiana.edu}

\author{Franz-Josef Zickgraf\altaffilmark{3}}
\affil{Hamburger Sternwarte, Universit\"at Hamburg, Gojensbergsweg 112, D-21029 Hamburg,
Germany}
\email{fzickgraf@hs.uni-hamburg.de}

\author{Johan Holmberg\altaffilmark{3}}
\affil{Max Planck Institute for Astronomy, Koenigstuhl 17, DE-69117 Heidelberg,
Germany}
\email{holmberg@mpia.de}

\author{Anna \"Onehag\altaffilmark{3}, Anders Eriksson\altaffilmark{3}}
\affil {Department of Astronomy and Space Physics, University of Uppsala 
Box 515, SE 751 20 Uppsala, Sweden}
\email{annao@astro.uu.se, anderse@astro.uu.se}

\author{Donald M. Terndrup}
\affil{Department of Astronomy, Ohio State University,
140 W. 18th Avenue, Columbus, OH 43210}
\email{terndrup@astronomy.ohio-state.edu}

\author{Samir Salim}
\affil{Department of Physics and Astronomy, University of California at Los
Angeles, Los Angeles, CA 90095}
\email{samir@astro.ucla.edu}

\author{Johannes Andersen}
\affil{The Niels Bohr Institute, Astronomy, Juliane Maries Vej 30,
DK-2100 Copenhagen, Denmark, and Nordic Optical Telescope Scientific
Association, Apartado 474, ES-38 700 Santa Cruz de La Palma, Spain}
\email {ja@astro.ku.dk}

\author{Birgitta Nordstr\"om}
\affil{The Niels Bohr Institute, Astronomy, Juliane Maries Vej 30,
DK-2100 Copenhagen, Denmark, and Lund Observatory, Box 43, S-221 00 Lund,
Sweden}
\email {birgitta@astro.lu.se}

\author{Norbert Christlieb}
\affil{Hamburger Sternwarte, Universit\"at Hamburg, Gojensbergsweg 112, D-21029 Hamburg,
Germany, and Department of Astronomy and Space Physics, Uppsala University,
Box 515, SE-75120, Uppsala, Sweden}
\email{norbert@astro.uu.se}

\author{Anna Frebel}
\affil{Research School of Astronomy \& Astrophysics, Australian National University, 
Cotter Road, Weston, ACT 2611, Australia}
\email{anna@mso.anu.edu.au}

\altaffiltext{1}{Visiting Astronomer, European Southern Observatory}
\altaffiltext{2}{Visiting Astronomer, WIYN 0.9m.  The 0.9m telescope is
operated by WIYN Inc. on behalf of a Consortium of ten partner Universities and
Organizations (see http://www.noao.edu/0.9m/general.html). WIYN is a joint
partnership of the University of Wisconsin at Madison, Indiana University, Yale
University, and the National Optical Astronomical Observatory.}
\altaffiltext{3}{Visiting Astronomer, Danish 1.5m Telescope}

\begin{abstract}

We report broadband $UBV$ and/or $BVR_CI_C$ CCD photometry for a total of 1857
stars in the thick-disk and halo populations of the Galaxy. The majority of
our targets were selected as candidate field horizontal-branch or other A-type
stars (FHB/A, N = 576), or candidate low-metallicity stars (N = 1221), from
the HK and Hamburg/ESO objective-prism surveys. Similar data for a small
number of additional stars from other samples are also reported.

These data are being used for several purposes. In the case of the FHB/A
candidates they are used to accurately separate the lower-gravity FHB stars from
various higher-gravity A-type stars, a subsample that includes the so-called
Blue Metal Poor stars, halo and thick-disk blue stragglers, main-sequence A-type
dwarfs, and Am and Ap stars. These data are also being used to derive
photometric distance estimates to high-velocity hydrogen clouds in the Galaxy
and for improved measurements of the mass of the Galaxy. Photometric data for
the metal-poor candidates are being used to refine estimates of stellar metallicity
for objects with available medium-resolution spectroscopy, to obtain distance
estimates for kinematic analyses, and to establish initial estimates of
effective temperature for analysis of high-resolution spectroscopy of the stars
for which this information now exists.

\end{abstract}


\section{Introduction}

Over the course of the past few decades two large objective-prism
surveys have enormously expanded the numbers of recognized members of the
thick-disk and halo populations of the Galaxy, the HK Survey of Beers and
colleagues (Beers, Preston, \& Shectman 1985, 1992a; Beers 1999), and the
Hamburg/ESO Survey (HES: Wisotzki et al. 1996; Christlieb 2003). 

As a result of their great numbers and high luminosities the field
horizontal-branch (FHB) stars in these surveys are nearly ideal tracers of the
kinematics and dynamics throughout the inner halo ($R < 20$ kpc) of the Galaxy
(e.g., Sommer-Larsen et al. 1997; Wilhelm et al. 1999b). The nearby FHB stars
are of particular importance because many of them either already have proper
motions available (e.g., UCAC2; Zacharias et al. 2004; USNOB+SDSS; Munn et al.
2004), or they will be measured in the near future (e.g., from ongoing surveys such as
the Southern Proper Motion program of van Altena and colleagues, see Girard et al.
2004). When combined with radial velocity information these data allow for
refined estimates of the mass of the Galaxy (e.g., Sakamoto, Chiba, \& Beers
2003). Also of great interest are the very metal-poor stars from these surveys
(stars with metallicities [Fe/H] $\le -2.0$, according to the nomenclature of
Beers \& Christlieb 2005), as they provide crucial elemental abundance
information needed to elucidate the nature of the first generations of stars to
form in the Galaxy. 

Photometric surveys, such as the ongoing Century Survey Halo project (Brown al.
2003), searches of the 2MASS Point Source Catalog (Brown et al. 2004), and the
Sloan Digital Sky Survey (SDSS: Gunn et al. 1998; York et al. 2000; Yanny et al.
2000; Sirko et al. 2004a,b; Clewley et al. 2005) are providing samples of
FHB stars with distances from 5 to more than 100 kpc from the Galactic center.
The extension of the SDSS, known as SDSS-II, includes the program SEGUE: Sloan
Extension for Galactic Understanding and Exploration. SEGUE will greatly enlarge
the list of known distant FHB stars. Given the large range of distances probed
by the FHB stars, they provide excellent tools for dynamical determination of
the mass distribution of the inner and outer halo (e.g., Sakamoto et al. 2003
and references therein; see also Sirko et al. 2004b), searches for streams of
disrupted dwarf galaxies (e.g., Yanny et al. 2000; Newberg et al. 2002; Yanny et
al. 2003; Belokurov et al. 2006), and for bracketing estimates of distance to
the high-velocity clouds of hydrogen in the Galaxy (e.g., Wakker et
al. 1996; Wakker \& van Woerden 1997; Wakker 2004; Thom et al. 2006).
 
Broadband photometry for metal-poor (MP) candidates provides information
required to obtain improved metallicities for these stars (in combination with
medium-resolution spectroscopy), as well as for the selection of appropriate
model atmospheres for analysis of high-resolution spectroscopy (e.g., Cayrel et
al. 2004). Accurate distance estimates for the MP stars are needed in order to
make full use of available proper motion information to carry out detailed
kinematic studies (e.g., Chiba \& Beers 2000).

In the color range $-0.2 \le B-V \le 0.4$, FHB stars have medium-resolution
spectra that are quite similar to other A-type stars of higher surface gravity.
When analysis of these spectra is combined with measured $UBV$ colors they
can be reliably distinguished (e.g., Wilhelm, Beers, \& Gray 1999a). Although it
was once thought that the numbers of FHB stars in this color range were likely
to dominate the higher surface-gravity stars (e.g., blue stragglers) in the halo
of the Galaxy, recent studies have shown that color-selected samples may
comprise up to 50\% from each of these populations (Norris \& Hawkins 1991;
Preston, Beers, \& Shectman 1994; Wilhelm et al. 1999b; Brown et al. 2004).
Finally, we recall that metal-poor main-sequence turnoff (TO) and subgiant (SG)
stars in the color range $0.3 \le B-V \le 0.4$ can be distinguished from FHB
stars on the basis of their generally higher surface gravities, and its effect
on the $U-B$ color (e.g., Wilhelm et al. 1999a; Bonifacio, Monai, \& Beers 2000).

Refinement of the classifications of FHB/A stars is necessary in order
to optimize their utility as probes of the Galaxy. The published catalogs of
FHB/A candidates from the HK survey (Beers, Preston, \& Shectman 1988; Beers et
al. 1996; Beers et al. 2006) are certain to comprise a confounded sample, as
representatives of both high- and low-surface gravity stars (as well as outright
errors due to mis-classification of prism spectra) are present. The HES contains
over 6800 stars that are classifiable as likely FHB stars, some as faint as $B =
17.5$. One of our primary reasons for undertaking the photometric study of the
FHB/A stars was to test a new method for distinguishing FHB stars from their
high-surface gravity counterparts on the basis of the HES objective-prism
spectra alone. This new approach makes use of a ``Str\"omgren Index'' based on
indices obtained directly from the HES prism spectra. Details of this approach,
and the results of our tests (using data from the present paper), have been
discussed in a separate paper (Christlieb et al. 2005).

In \S 2 we discuss the selection of stars for inclusion in the present study. 
The observations and data reduction procedures are presented in \S 3.  The
catalog of $UBVR_CI_C$ photometry is described in \S 4.  For convenience of later
use, this information is supplemented by near-infrared $JHK$ photometry provided
by matches to the 2MASS Point Source Catalog (Skrutskie et al. 2006).  In \S 5
we present a brief discussion of the stars contained in the present catalog.
 
\section{Sample Selection}

Stars for this study were selected for a variety of reasons.

The majority of our candidate FHB/A stars were selected from the HK survey and
the HES. A few stars were selected from other sources. These include MP
candidates from the HK-II survey of Rhee (2000, 2001), FHB/A candidates from
2MASS, and suspected RR Lyraes from the Robotic Optical Transient Search
Experiment (ROTSE-I) Northern Sky Variability Survey (NSVS) catalog (Wozniak
2004), which are being used for an extensive follow-up effort to estimate
distances to high-velocity clouds. 

The $UBV$ data we have obtained has been used, in conjunction
with available medium-resolution spectroscopy, to assess our ability to uniquely
separate the low-gravity FHB stars from the generally higher surface gravity
blue straggler and main-sequence A-type stars (Christlieb et al. 2005). The
selection of 125 test objects for which we report $UBV$ photometry in the
present paper is described fully by Christlieb et al. (2005). In short, these
objects were chosen to cover a range of the approximate Str\"omgren indices
measured from the HES prism plates.  These were then used to evaluate the relative
success of gravity separation by this approach, by checking their derived
classifications with those obtained from the techniques of Wilhelm et al.
(1999a). Christlieb et al. (2005) demonstrated that the maximum contamination in
samples of HES FHB candidates by higher-gravity A-type stars, when classified via the
Str\"omgren approach, is no more than about 16\%.

Our present catalog includes $UBV$ photometry for an additional 451 FHB/A candidates
that are being used to expand the work of Sommer-Larsen et al. (1997), in order
to study the proposed change in the nature of the velocity ellipsoid from the
inner to outer halo of the Galaxy. Thom et al. (2005) presents the first results
from this ongoing effort.  

Some 500 candidate MP stars in our catalog were included because they now have
high-resolution spectroscopy available from follow-up studies with 8m-10m class
telescopes. Among these programs are the ``First Stars'' study of Cayrel et al.
(2004) with VLT/UVES, and studies of extremely metal-poor stars conducted with
Subaru/HDS (e.g., Aoki et al. 2002, 2005, 2006; Honda et al. 2004a,b), and the
``0Z'' project of Cohen and collaborators (Carretta et al. 2002; Cohen et al.
2002, 2004). A very large sample of high-resolution spectroscopy for over 350
candidate very metal-poor giants with [Fe/H] $\le -2.0$, selected primarily from
the HES, has recently been reported on by Christlieb et al. (2004) and Barklem
et al. (2005). This program, known as the Hamburg/ESO R-process Enhanced Star
(HERES) survey, used VLT/UVES to obtain moderate signal-to-noise ($S/N \sim
30/1-50/1$ per pixel), $R \sim 20000$ spectra in order to identify additional
examples of highly r-process-element enhanced stars similar to the well-known
stars CS~22892-052 (Sneden et al. 2003) and CS~31082-001 (Cayrel et al. 2001;
Hill et al. 2002), as well as to better assess their frequency of occurance over
this metallicity range. Many of these stars were included in our sample
selection. A number of the bright metal-poor stars selected by Frebel et al.
(2006) for which high-resolution spectroscopy either presently exists, or
will soon be obtained, were also included in our observational program.

Numerous stars from the HK and HES surveys exhibit strong CH G-bands, indicative
of large carbon abundances (see, e.g., Christlieb et al. 2001; Marsteller et al.
2005; Lucatello et al. 2006). Many (roughly 100) have had high-resolution
spectroscopy gathered by a number of groups (e.g., Barklem et al. 2005,
Tsangarides et al. 2005; Aoki et al. 2006); these stars were included in our
observing program as well. 

\section{Observations and Data Reduction}

The photometry observations reported in the present paper were conducted over a
total of 28 runs with 6 different telescope/detector combinations during the
period 1998 to 2006. There were 13 runs with the ESO/Danish 1.5m telescope on La
Silla, 10 runs with the WIYN 0.9m telescope on Kitt Peak, three runs with the
CTIO 0.9m (including two conducted by the SMARTS consortium), one run with the
CTIO 1.0m (conducted by the SMARTS consortium), and a single run with the
Hiltner 2.4m telescope at the MDM observatory on Kitt Peak. Details of the dates
of the runs, and the observers who participated in each run, are listed in Table
1.

\subsection{The ESO/Danish 1.5m Observations}

Observations were obtained with the ESO/Danish 1.5m telescope at the European
Southern Observatory using the DFOSC instrument, as described by Brewer \& Storm
(1999). Prior to September 2000, the C1W7 CCD (a Loral/Lesser
backside-illuminated chip with 15 $\mu$m pixels) was employed. After this date a
new EEV CCD was installed, with the same pixel size, an improved readout noise (
$3e^-$ vs. 7e$^-$ for the previous chip), and a much higher full-well capacity
than the previous chip. In both cases, the ESO $UBV$ and/or $BVR_CI_C$ filter
sets were employed. 

A reasonably clean and uniformly illuminated portion of the full chip was
identified, and the program stars were located close to its center. Because we
were primarily observing single stars in each pointing, only the central 250
$\times$ 250 pixels surrounding the region of interest were read out.
Observations were typically taken by running sequences of $V-B-U-U-B-V$ or
$V-B-R_C-I_C-I_C-R_C-B-V$ filters, with a bias taken after each cycle.

The goal was to set integration times so as to obtain a minimum of 15,000 net
counts above sky in each filter, so that the summed observations were at least
30,000 net counts. In the case of particularly faint stars the sequences were
repeated until the desired sum was obtained, as best as could be judged at the
telescope. This goal was naturally much easier met for the brighter stars. In
the case of the fainter stars, required integration times became sufficiently
long that one must be concerned about changing sky brightness during the course
of the observations. In most all cases the longest individual integration times
were set to 20 minutes. As a result (and as reflected in Table 2), the
photometric errors for fainter stars grow with increasing apparent magnitude. 

Nightsky flatfields were obtained (when possible) at the beginning and end of
each night, using the same 250 $\times$ 250 pixel region as used for the program
stars. In later runs, full-chip flatfields were obtained for use in flatfielding
the generally much larger fields containing the standards, so that more
standards could be used. The standards we employed were taken from the list of
Landolt (1992), supplemented in some cases by the Graham (1982) E-region
standards. At least several times (typically 3 to 5 times) per night,
observations of 2 or 3 extinction standards were attempted (weather permitting),
covering a range in airmass from 1.0 to 2.5.  These stars were used to provide
an independent check on the airmass terms in the final photometric solutions.

\subsection{The WIYN 0.9m Observations}

Observations were obtained with the WIYN 0.9m telescope on Kitt Peak, using the
S2KB CCD mounted at the Cassegrain focus. The detector was a SITe 2048 $\times$
2048 pixel array with a plate scale of 0.6\arcsec/pixel, yielding a field of
view of 20.5\arcmin{} $\times$ 20.5\arcmin. We employed the standard Harris $UBV$ and/or
$BVR_CI_C$ filter sets, which well reproduce the properties of the Johnson $UBV$
and Kron-Cousins $R_CI_C$ system.  

Dome flatfields were obtained during each afternoon preceding the night of
observations. Standards were taken from the equatorial list of Landolt (1992).  For
program stars, a cosmetically clean section of the chip covering 200 $\times$ 200
pixels was used.  Standards were in general taken using the full chip.  The
sequence of filters, and the targeted net counts (30,000) were obtained in a
manner identical to the procedure employed with the ESO/Danish 1.5m described
above. 

\subsection{The CTIO 0.9m / Yale 1.0m Observations}

Broadband $UBV$ and $BVR_CI_C$ observations were taken with the CTIO 0.9m and
Yale 1.0m telescopes. Two runs were obtained during SMARTS queue-mode operation,
while one run was conducted in classical mode. In all cases, the detector was a
2048 $\times$ 2046 QUAD amplifier CCD with a 13.6\arcmin{} field of view. The CCD
has a pixel scale of 0.396\arcsec/pixel. The CCD was operated in the QUAD
amplifier mode for all runs.  

Dome flatfields were obtained in the afternoon before each observing night.
In addition, sky flatfields were taken in the evening to supplement the dome
flatfields.

Photometric standard fields were taken from the Landolt (1992) equatorial
standard catalog. Standards were taken throughout the nights so
that we could characterize airmass extinction coefficients and changes in
atmospheric conditions. Each program star was observed using sequences such as
those for the ESO/Danish 1.5m described above. The program stars were
always positioned in a cosmetically clean portion of the CCD, using a single
amplifier.  

\subsection{The Hiltner 2.4m Observations}

$BVR_CI_C$ Observations were obtained with the 2.4m Hiltner Telescope of the MDM
Observatory. The detector was the ``Echelle'' CCD, with a frame scale of 0.28
\arcsec/pixel. The central 512 $\times$ 512 pixel region of the chip was used.
Nightly sky flats were obtained in each filter.   

Photometric standards, taken from the Landolt (1992) equatorial lists, were
observed on each night. Atmospheric extinction terms were found for each night;
the instrumental color terms were taken as the average throughout the run. For
both the standards and the program stars, the normal practice was to take three
exposures in $BVR_CI_C$ at each pointing.

\subsection {Reductions}

The data were reduced using the photometry packages found within
IRAF\footnote{IRAF is distributed by the National Optical Astronomy
Observatories, which is operated by the Association of Universities for Research
in Astronomy, Inc. under cooperative agreement with the National\ Science
Foundation.}. We employed the packages {\tt ccdproc}, {\tt zerocombine}, and
{\tt flatcombine} in order to trim the images and assemble a master bias that
was applied to all of the images taken on a given night of each run. Master
flats in each filter were then constructed for each filter and applied to the
appropriate target fields. The package {\tt phot} was then used to obtain
aperture photometry for all of the standard and target stars. In this procedure,
a single (typically a 15\arcsec{} aperture) was defined as the region including the light
of the star, and an annulus of 5\arcsec{} width, located 5\arcsec{} outside the region of the
stellar aperture, was used to estimate the local sky background, which was
scaled appropriately and subtracted from the flux included in the stellar
aperture.

We then employed the packages {\tt mknobsfile}, {\tt mkconfig}, {\tt fitparams}
and {\tt invertfit} to obtain the instrumental magnitudes of the standard stars;
these were then used to determine the coefficients for the transformation
equations for each magnitude or color. These solutions were then applied to the
standard stars to verify their validity before applying them to the target stars
to obtain calibrated magnitudes. As a quality check, stars from the various
programs were observed on multiple nights during each run, and also across the
various telescope/detector combinations. A number of stars were identified from
this procedure for which the data appeared less than optimal; these stars are
noted in the reported results.

\section{Results}

Although the quality of the photometry obtained varied from telescope to
telescope, and from run to run, we have a sufficient number of independently
observed stars taken during the program to estimate the scatter obtained over
the entire sample. These are listed, as a function of $V$ magnitude ranges,
in Table 2. This Table lists the average one-sigma errors in the measured $V$
magnitudes and in the optical colors obtained from our observations and
reductions. These estimates provide a good picture of the level of (internal)
accuracy one can expect for the great majority of stars reported here. The total
number of observations involved are listed in the second column of Table 2. The
average internal errors are listed in the third column of Table 2, on the lines
labeled ``INT''. These errors are dominated by photon statistics and residuals
in the reduction solutions. As can be seen from inspection of the Table, the
typical internal errors range from 0.002 mag to 0.006 mag for the brightest
stars, to on the order of 0.02 mag to 0.05 mag for the faintest stars.

In order to obtain an estimate of the typical external errors in our
measurements, we have also computed the average one-sigma errors in the measured
$V$ magnitudes and optical colors for stars that were observed across multiple
runs, and/or with multiple telescope/detector combinations. The total number of
observations involved are listed in the second column of Table 2. The average
external errors are listed in the third column of Table 2, on the lines labeled
``EXT''. As can be seen from inspection of the Table, the typical external
errors range from 0.010 mag to 0.012 mag for the brightest stars, to on the order
of 0.027 mag to 0.037 mag for the faintest stars. We conclude that for stars
brighter than $V = 15$, the external errors of determination in the apparent
magnitudes and colors are on the order of 2\%, while errors of 3\% to 3.5\%
apply for the stars fainter than $V = 15$.

We observed a sufficient number of stars that had previously reported $UBV$
photometry (see, e.g., Pier 1982; Doinidis \& Beers 1990, 1991; Preston,
Shectman, \& Beers 1991; Beers et al. 1992b; Preston et al. 1994; Norris, Ryan,
\& Beers 1999; Wilhelm et al. 1999b; Bonifacio et al. 2000), obtained during the
course of the HK survey, to obtain an independent comparison. Figure 1 shows a
plot of the comparison of the $V$ magnitudes, $B-V$, and $U-B$ colors for the
stars in common. As can be seen in the Figure, the agreement is excellent. From
the 185 stars in common between the present and previous samples, we obtain
average zero-point offsets of:

\begin{eqnarray*}
< V_{\rm pres} - V_{\rm prev} >         & = & -0.005 \; {\rm mag}\\
< (B-V)_{\rm pres} - (B-V)_{\rm prev} > & = & +0.005 \; {\rm mag}\\
< (U-B)_{\rm pres} - (U-B)_{\rm prev} > & = & -0.001 \; {\rm mag}\\
\end{eqnarray*}

\noindent The average one-sigma scatter in the different measurements is:

\begin{eqnarray*}
<\sigma \; ( V_{\rm pres} - V_{\rm prev} ) >        & = & 0.026 \; {\rm mag}\\
<\sigma \; ( (B-V)_{\rm pres} - (B-V)_{\rm prev} )> & = & 0.028 \; {\rm mag}\\
<\sigma \; ( (U-B)_{\rm pres} - (U-B)_{\rm prev} )> & = & 0.022 \; {\rm mag}\\
\end{eqnarray*}

\noindent  In the above expressions ``pres'' refers to the present
sample and ``prev'' refers to the previous samples. 

The great majority of the stars in common have $V \le 15$; the average one-sigma
scatter between the two sets of stars are commensurate with the expected level 
of external errors listed in Table 2.  There were not a large enough number of
stars observed across the various telescope/detector combinations to obtain a
breakdown of how the zero points and scatter might vary for each of these
combinations relative to previously reported photometry. However, sanity checks
performed during the course of assembling the results for the small number of
previously observed stars did not indicate cause for concern.

Table 3 lists the final results for our program stars. For each star we list
the star name, the $V$ magnitude, the $B-V$, $U-B$, $V-R_C$, and $V-I_C$ colors,
along with the one-sigma errors of their determination, and the total number of
independent observations obtained for each magnitude or color. Stars for which
our repeated measurements indicated that the reported results might be suspect
are noted in the Table with a ``:'' following the reported photometry. This
includes a number of stars with extremely red colors (e.g., $B-V > 2.5$), for
which in most cases there was a lack of sufficiently red standards observed
during the course of the run to be confident of the resulting solutions.

We have searched the 2MASS Point Source Catalog (Skrutskie et al. 2006) in order
to obtain $JHK$ magnitudes for our program sample. Table 4 summarizes the
results of this exercise. This Table lists the star name, equinox 2000
coordinates, Galactic longitude and latitude, the estimated line-of-site
reddening (described below), the 2MASS identification, and the 2MASS magnitudes
and associated errors. Note that we were not able to match up some of our
program stars, either because they were too faint to be included in the 2MASS
catalog, or because the coordinates from the HK survey were not sufficiently
accurate to enable a confident identification. Stars for which the 2MASS catalog
match indicated that there were possible problems with the reported photometry
(due, for instance, to contamination from nearby stars, or other difficulties)
are indicated in the Table with a ``:'' following the reported 2MASS photometry.  

The estimated line-of-sight reddening toward each star is obtained from the dust
maps of Schlegel, Finkbeiner, \& Davis (1998), which have superior spatial
resolution and are thought to have a better-determined zero point than the
Burstein \& Heiles (1982) maps. In cases where the Schlegel et al. estimate
exceeds $E(B-V)_S$ = 0.10 we follow the procedures outlined by Bonifacio et al.
(2000) to reduce these estimates by 35\%, and obtain the adopted
estimate of reddening, $E\;(B-V)_A$.

Tables 3 and 4 are split into five primary categories: (1) stars from the HK
survey, (2) stars from the HES, (3) stars from the HK-II sample of Rhee
(2000, 2001) and colleagues, (4) stars from the ongoing HVC tracer star sample of
York and colleagues, and (5) stars from other sources. Each star is listed in
these sections, ordered by its name. The HK and HK-II survey stars are grouped
according to their plate number. Table 5 provides a listing of stars we are
aware of in our listing that appear in several surveys (``Other Name'') or which
have several names (due to overlapping plates) in the HK survey (``Repeat
Object'').

\section {Discussion}

Our catalog includes broadband $UBV$ and/or $BVR_CI_C$ photometry for a total
of 1857 stars, obtained over the course of 28 runs conducted at various
telescopes over the past 8 years. The majority of the stars in this catalog (N =
1221) were selected as candidate metal-poor candidates from either the HK survey
or the Hamburg/ESO survey.  The other large category of targets were the
candidate FHB/A stars selected from these same surveys (N = 576).  A smaller
number of additional stars were selected either as metal-poor, FHB/A, or RR Lyrae
candidates from other surveys, as described in \S 2 above.

Figure 2a is a histogram of the distribution of $V$ magnitude for the stars
included in this study (not including stars where the photometry was suspect).
Figures 2b-2e are similar histograms that indicate the distribution of measured
$B-V$, $U-B$, $V-R_C$, and $V-I_C$ for the stars where this information is
available. Figures 2f-2h are histograms of the 2MASS $J-K$, $H-K$, and $J-K$
colors for objects that we were able to match to our program sample.

In Figure 3 we show a de-reddened two-color diagram, $(U-B)_0$ vs. $(B-V)_0$,
for the entire set of candidates with that information available. Numerous
low-metallicity TO stars can be seen in this Figure in the color range $0.3 \le
(B-V)_0 \le 0.5$, as they are displaced upward from the Population I
main-sequence relation. The red dashed line indicates the expected location of
main-sequence dwarfs with metallicities [Fe/H] $= -4.0$, according to synthetic
colors based on the Kurucz (1993) models.  Note that many of the stars with 
$(B-V)_0 > 0.5$ are likely to be giants. The solid green line indicates the
rough division between dwarfs that are expected to be on the main sequence and
those stars that may be considered likely Blue Metal Poor (BMP; see Preston et
al. 1994) or FHB stars. The BMP candidates are located in the region closest to
the red dashed line and blueward of the green line, extending to $(B-V)_0 \sim 0.15$.
Many of the stars in the color range $-0.2 \le (B-V)_0 \le
0.3$ are metal-poor FHB stars (see Beers et al. 1995; Bonifacio et al. 2000).
Note that the low surface gravities of FHB stars displace their positions in the
two-color diagram to locations that appear consistent with the Population I
dwarf line.  The bona-fide FHB stars among these candidates cover ranges of distance between
a few kpc up to 75 kpc from the Sun, depending on the color.

All of the stars in our program sample have medium-resolution ($\sim$ 2 \AA\ )
optical spectroscopy available. Many of the most interesting low-metallicity
stars in our sample either already have high-resolution spectroscopy available,
or will be studied at high-resolution in the near future. Studies of the
kinematics of the BMP/FHB stars included in this sample will be reported in due
course.

\acknowledgements

We are thankful to the ESO/Danish 1.5m, WIYN 0.9m, NOAO, and MDM time
assignment committees for awards of the significant amounts of telescope time
required for this project, and for the patience to await the results. We are
also grateful for the excellent support that we received at the telescopes.

T.C.B., Y.L., B.M., and N.D. acknowledge partial support from grants AST
00-98508, AST 00-98549, AST 04-06784, and PHY 02-16783, Physics Frontier
Centers/JINA: Joint Institute for Nuclear Astrophysics, awarded by the U.S.
National Science Foundation.

J.K. acknowledges partial support from the Honors College at Michigan State
University, in the form of a Professorial Assistantship.

C.F. acknowledges partial support from the Danish National Research Foundation
and the Carlberg Foundation.

S.R. acknowledges partial support from the Brazilian institutions
FAPESP, CNPq and Capes.

C.P.D. acknowledges partial support from grant AST 02-06202, awarded by the U.S.
National Science Foundation.

D.T. and S.S. acknowledge partial support from grant AST 02-05789, awarded by
the U.S. National Science Foundation, and from the Ohio State Program for the
Enhancement of Graduate Studies (PEGS).

J.A. and B.N. thank the Carlsberg Foundation and the Swedish and Danish 
Natural Science Research Councils for partial financial support of this work.

N.C. acknowledges partial support from the Deutsche Forschungsgemeinschaft through
grants Ch~214/3 and Re~353/44. N.C. is a research fellow of the Royal Swedish
Academy of Sciences supported by a grant from the Knut and Alice Wallenberg
Foundation. 

A.F. acknowledges partial support from grant DP0342613, awarded by the
Australian Research Council.

\clearpage

\begin{figure}
\epsscale{0.60}
\plotone{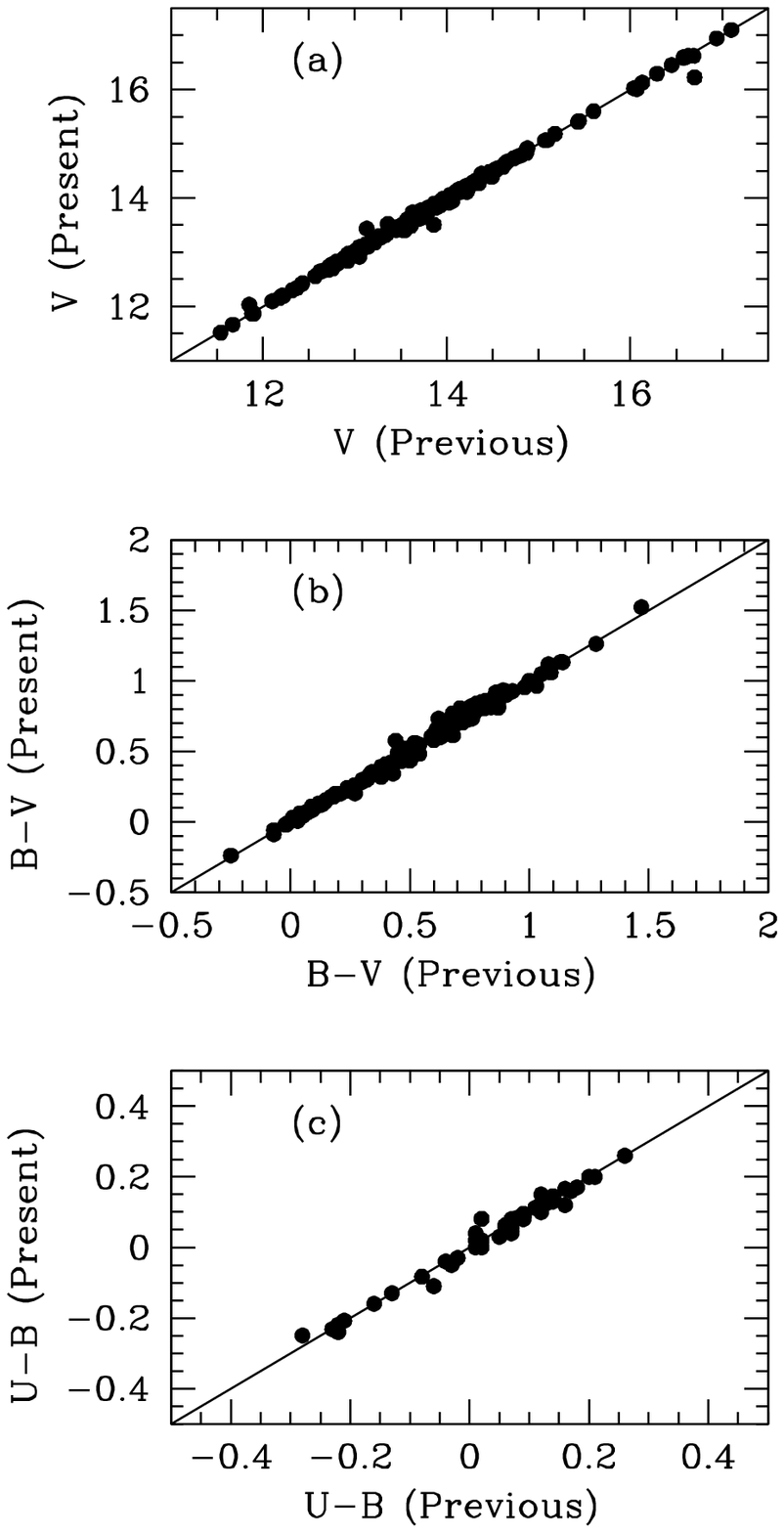}
\caption{Comparison of measured (a) $V$, (b) $B-V$, and (c) $U-B$ for stars in the present
paper in common with stars haveing previously reported photometry, from sources
listed in the text. The solid line is the one-to-one line. The agreement is
excellent.}
\end{figure}

\clearpage

\begin{figure}
\epsscale{1.0}
\plotone{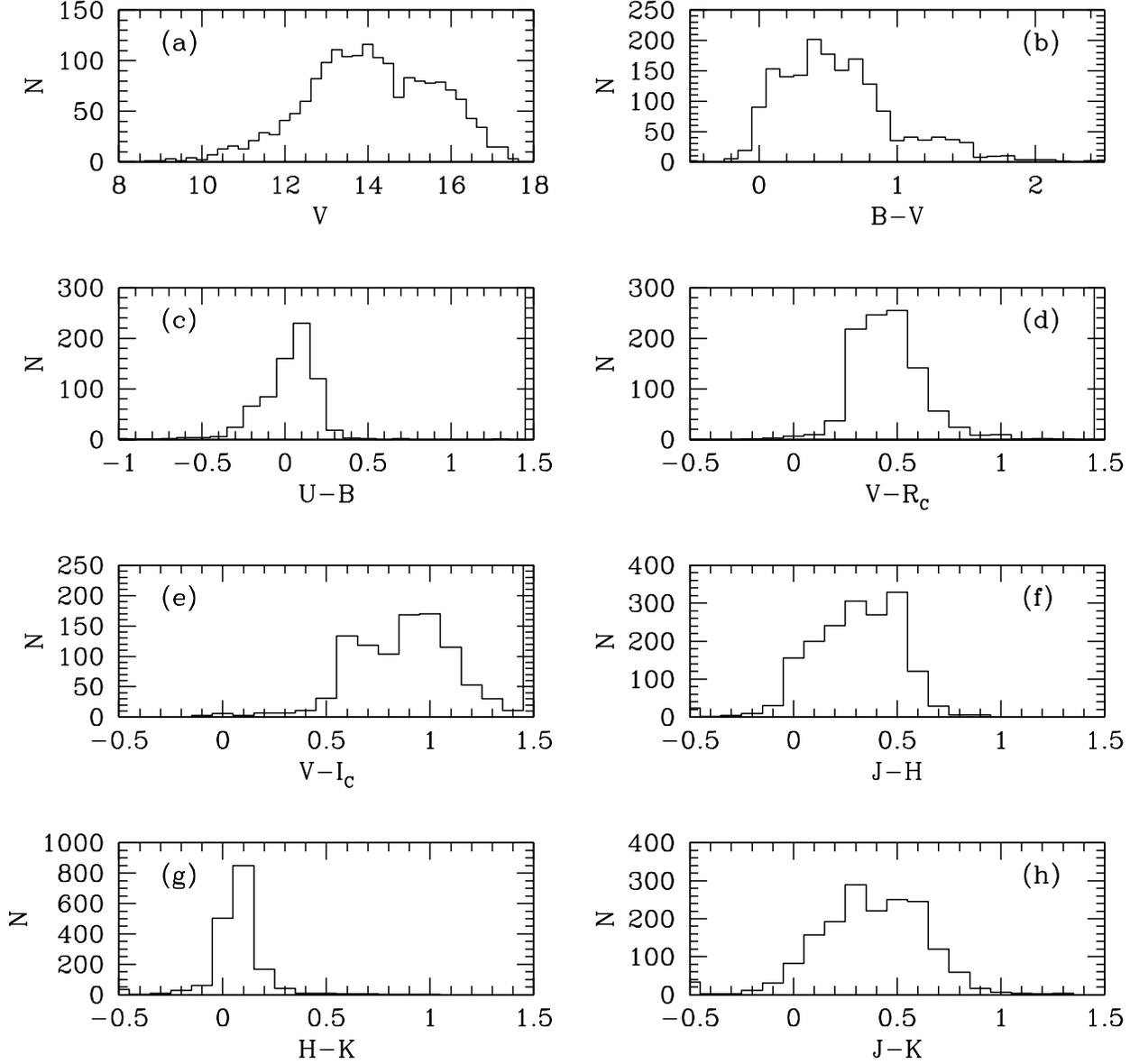}
\caption{(a) Distribution of $V$ magnitudes for stars in the present sample. 
Bins are 0.25 magnitudes in width. (b) Similar for $B-V$ colors. (c) Similar for
$U-B$ colors. (d) Similar for $V-R_C$ colors. (e) Similar for $V-I_C$ colors.
(f) Similar for $J-H$ colors. (g) Similar for $H-K$ colors. (h) Similar for
$J-K$ colors. Bins for the colors are 0.1 magnitudes in width. }
\end{figure}

\clearpage

\begin{figure}
\epsscale{0.60}
\plotone{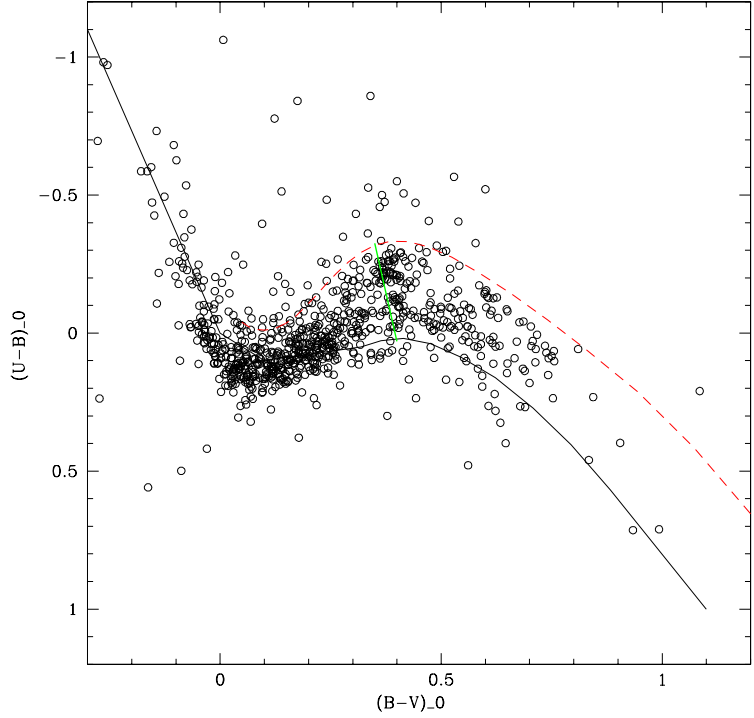}
\caption{De-reddened two-color $(U-B)_0$ vs. $(B-V)_0$ diagram for our program stars with
this information. The solid line is the Johnson (1966) Population I
main-sequence line. The red dashed line is obtained from the predicted $U-B$ and
$B-V$ colors from the Kurucz (1993) models for main-sequence stars with [Fe/H]
$= -4.0$. The solid green line represents the limit of the expected location of
the main-sequence TO for stars of the thick-disk and halo populations. The BMP
candidates are located in the region closest to the red dashed line and blueward
of the green line, extending to $(B-V)_0 \sim 0.15$. Many of the stars in the
color range $-0.2 \le (B-V)_0 \le 0.3$ are metal-poor FHB stars.
See text for additional comments.}
\end{figure}

\end{document}